\documentclass[preprint,eqsecnum,aps,nofootinbib]{revtex4}
\usepackage[dvips]{graphicx}
\usepackage{amssymb}
\usepackage{amsmath}
\usepackage{amsmath, amsthm, amssymb, mathrsfs}
\usepackage{pstricks}

\newcommand{\METER}{\psframe( 0.00,-0.25)( 0.50, 0.25)%
                    \psarc[linewidth=0.2pt]( 0.25,-0.20){0.25}{50}{130}%
                    \psline[linewidth=0.2pt, arrows=->]( 0.25,-0.20)( 0.35, 0.2)
           }

\begin{document}

\title{Mixed-State Entanglement and Quantum Teleportation through Noisy Channels}
\author{
Eylee Jung, Mi-Ra Hwang, DaeKil Park}

\affiliation{Department of Physics, Kyungnam University, Masan,
631-701, Korea}

\author{Jin-Woo Son}

\affiliation{Department of Mathematics, Kyungnam University, Masan,
631-701, Korea}

\author{S. Tamaryan}

\affiliation{Theory Department, Yerevan Physics Institute,
Yerevan-36, 375036, Armenia}

\vspace{1.0cm}

\begin{abstract}
The quantum teleportation with noisy EPR state is discussed. Using an optimal 
decomposition technique, we compute the concurrence, entanglement of formation and 
Groverian measure for various noisy EPR resources. It is shown analytically that all 
entanglement measures reduce to zero when $\bar{F} \leq 2/3$, where $\bar{F}$ is an 
average fidelity between Alice and Bob. This fact indicates that the entanglement is a 
genuine physical resource for the teleportation process. This fact gives valuable 
clues on the optimal decomposition for higher-qubit mixed states. As an example, the 
optimal decompositions for the three-qubit mixed states are discussed 
by adopting a teleportation with 
W-state.
\end{abstract}


\maketitle

\section{Introduction}

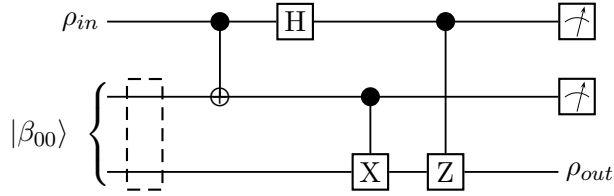
\begin{figure}
\begin{center}
\begin{pspicture}(-4,0)(4,3)
   \psframe(-0.75, 2.25)(-0.25, 2.75)\rput[c](-0.50, 2.50){H}
   \psframe( 0.25, 0.25)( 0.75, 0.75)\rput[c]( 0.50, 0.50){X}
   \psframe( 1.25, 0.25)( 1.75, 0.75)\rput[c]( 1.50, 0.50){Z}
   
   \rput[c]( 3.0, 2.50){\METER}
   \rput[c]( 3.0, 1.50){\METER}
   
   \psframe[linestyle=dashed](-2.75, 0.25)(-2.25, 1.75)

   \psline[linewidth=0.7pt](-3.00, 2.5)(-0.75, 2.5)
   \psline[linewidth=0.7pt](-0.25, 2.5)( 3.00, 2.5)
   
   \psline[linewidth=0.7pt](-3.00, 1.5)( 3.00, 1.5)
   
   \psline[linewidth=0.7pt](-3.00, 0.5)( 0.25, 0.5)
   \psline[linewidth=0.7pt]( 0.75, 0.5)( 1.25, 0.5)
   \psline[linewidth=0.7pt]( 1.75, 0.5)( 3.00, 0.5)
   
   \psline[linewidth=0.7pt](-1.50, 2.50)(-1.50, 1.50)
   \psline[linewidth=0.7pt]( 0.50, 1.50)( 0.50, 0.75)
   \psline[linewidth=0.7pt]( 1.50, 2.50)( 1.50, 0.75)
   
   \psdot[dotstyle=*, dotscale=2](-1.50, 2.50)
   \psdot[dotstyle=oplus, dotscale=2](-1.50, 1.50)
   \psdot[dotstyle=*, dotscale=2]( 0.50, 1.50)
   \psdot[dotstyle=*, dotscale=2]( 1.50, 2.50)
   
   \rput[r](-3.1, 2.5){$\rho_{in}$}
   \rput[l]( 3.1, 0.5){$\rho_{out}$}
   \rput[c](-3.9, 1.0){$\lvert \beta_{00} \rangle$}
   \rput[c](-3.0, 1.0){$\left\{ \begin{array}{r} \\ \\ \end{array}  \right.$}

\end{pspicture}
\end{center}
\caption{A quantum circuit for quantum teleportation through noisy channels with
EPR state. The top two lines belong to Alice while the bottom line belongs to Bob.
The dotted box represents noisy channels, which makes the EPR state to be mixed
state.}
\end{figure}

Entanglement of quantum states plays a crucial role in modern quantum information 
theories\cite{nielsen00}. Although we do not have a general theory of the quantum 
entanglement, many physicists believe that it is a physical resource which makes 
quantum computer outperforms classical ones\cite{vidal03-1}. Thus in order to quantify
the entanglement of given quantum state many entanglement measures were constructed
during last decade. The basic entanglement measure is the entanglement 
of formation\cite{form1,form2,form3,form4}. Generally, entanglement of formation is 
defined in any bipartite system. For pure state if $|\psi\rangle$ is the state of the
whole system, the entanglement of formation ${\cal E}(\psi)$ is defined as von Neumann
entropy ${\cal E}(\psi) = - \mbox{Tr} \rho \log_2 \rho$, where $\rho$ is the 
partial trace over either of the two subsystems. Another measure we would like to 
use in this paper is Groverian measure\cite{biham01-1}. Groverian measure $G(\psi)$
for given $n$-qubit quantum state $|\psi\rangle$ is defined using a quantity
\begin{equation}
\label{pmax1}
P_{max} (\psi) = \max_{|q_1\rangle, \cdots, |q_n\rangle}
|\langle q_1| \cdots \langle q_n | \psi \rangle |^2
\end{equation}
where $|q_i\rangle$'s are single-qubit states. In fact, $P_{max} (\psi)$ is the 
maximal probability of success in the Grover's search algorithm\cite{grover97} when 
$|\psi\rangle$ is used as an initial state. Roughly speaking, $P_{max}$ quantifies a 
distance between a given $n$-qubit state $|\psi\rangle$ and a set of product states. 
Therefore, the entanglement should decrease with increasing $P_{max}$. In this 
reason Groverian measure is defined as $G(\psi) = \sqrt{1 - P_{max} (\psi)}$. For 
$2$-qubit pure states $P_{max}$ can be analytically computed\cite{shim04}, whose 
expression is 
\begin{equation}
\label{pmax2}
P_{max}  = \frac{1}{2} \left[1 + \sqrt{1 - 4 \mbox{det} \rho} \right]
\end{equation}
where $\rho$ is the partial trace over either of the two-qubits. Recently, $P_{max}$
for some $3$-qubit states were also computed 
analytically\cite{tama07-1,tama08-1,jung08-1} by exploiting a theorem of 
Ref.\cite{jung07-1}. Although much progress was developed recently for understanding 
the general features of pure-state entanglement, it seems to be far from complete
understanding.

The purpose of this paper is to examine the physical role of mixed-state entanglement.
In order to address this issue it is convenient to consider the quantum 
teleportation\cite{bennett93} when the quantum channel is affected by noise. The effect 
of noise in teleportation was discussed in Ref.\cite{oh02}. In order to explain the 
motivation of this paper it had better review Ref.\cite{oh02} briefly. Let us 
consider the usual situation of the teleportation: Alice and Bob share an EPR
channel
\begin{equation}
\label{EPR}
|\beta_{00} \rangle = \frac{1}{\sqrt{2}} ( |00\rangle + |11\rangle )
\end{equation}
and Alice wants to send a single-qubit state
\begin{equation}
\label{input}
|\psi_{in}\rangle = \cos \left( \frac{\theta}{2} \right) e^{i \phi/2} |0\rangle +
                    \sin \left( \frac{\theta}{2} \right) e^{-i \phi/2} |1\rangle.
\end{equation}
to Bob. We assume, however, that the perfect EPR state was not prepared initially 
due to noise. In terms of density operator language this means that instead of 
$\rho_{EPR} = |\beta_{00}\rangle \langle \beta_{00}|$ the imperfect density
operator $\varepsilon(\rho_{EPR})$ was made initially, where $\varepsilon$ is a quantum
operation. Since $\varepsilon(\rho_{EPR}) \neq \rho_{EPR}$ generally, Alice cannot send 
$|\psi_{in}\rangle$ perfectly to her remote recipient. This situation is depicted 
in Fig. 1. In this figure the top two lines belong to Alice while the bottom line 
belongs to Bob. The density operator $\rho_{in}$ is 
$|\psi_{in}\rangle \langle \psi_{in}|$ and $\rho_{out}$ is a state Bob receieves from 
Alice. The dotted box represents an imperfect EPR resource produced initially due to
the noise.

Two questions naturally arise at this stage. First one is what the explicit expression
of $\varepsilon(\rho_{EPR})$ is. Second one is how much information Alice can send 
to Bob. Obviously the answers are dependent on what type of noise we take into account.
To address the first question authors in Ref.\cite{oh02} used a master equation
in the Lindbald form\cite{lindbald76}
\begin{equation}
\label{lindbald}
\frac{\partial \sigma}{\partial t} = -i [H_S, \sigma] + \sum_{i, \alpha}
\left(L_{i,\alpha} \sigma L_{i,\alpha}^{\dagger} - \frac{1}{2}
\left\{ L_{i,\alpha}^{\dagger} L_{i,\alpha}, \sigma \right\} \right)
\end{equation}
where $\sigma \equiv \varepsilon(\rho_{EPR})$ and $L_{i,\alpha}$ is an Lindbald operators
which represent the type of noise. In order to simplify the situation Ref.\cite{oh02}
choosed simple types of noise $L_{i,\alpha} \equiv \sqrt{\kappa} \sigma^{(i)}_{\alpha}$ 
which acts on the
$i$th qubit to describe decoherence, where $\sigma^{(i)}_{\alpha}$ denotes the
Pauli matrix of the $i$th qubit with $\alpha = x,y,z$. The constant $\kappa$
is approximately equals to the inverse of decoherence time. The master equation
approach is shown to be equivalent to the usual quantum operation approach
for the description of noise in open quantum system\cite{nielsen00}.
Solving a master equation (\ref{lindbald}), we can now derive $\varepsilon(\rho_{EPR})$
explicitly. If we choose noises with same direction, {\it i.e.} $(L_{2,x},L_{3,x})$,
$(L_{2,y},L_{3,y})$, or $(L_{2,z},L_{3,z})$, Eq.(\ref{lindbald}) provides
\begin{eqnarray}
\label{mixed-xyz}
& &
\varepsilon_x(\rho_{EPR}) = \frac{1}{2}
\left(        \begin{array}{cccc}
       \tau_+  &  0  &  0  &  \tau_+          \\
       0  &  \tau_-  &  \tau_-  &  0          \\
       0  &  \tau_-  &  \tau_-  &  0          \\
       \tau_+  &  0  &  0  &  \tau_+
              \end{array}               \right) 
\hspace{1.0cm}
\varepsilon_y(\rho_{EPR}) = \frac{1}{2}
\left(        \begin{array}{cccc}
       \tau_+  &  0  &  0  &  \tau_+          \\
       0  &  \tau_-  &  -\tau_-  &  0          \\
       0  &  -\tau_-  &  \tau_-  &  0          \\
       \tau_+  &  0  &  0  &  \tau_+
              \end{array}               \right)
                                                      \\   \nonumber
& & \hspace{2.5cm}
\varepsilon_z(\rho_{EPR}) = \frac{1}{2}
\left(        \begin{array}{cccc}
       1  &  0  &  0  &  e^{-4 \kappa t}          \\
       0  &  0  &  0  &  0          \\
       0  &  0  &  0  &  0          \\
       e^{-4 \kappa t}  &  0  &  0  & 1 
              \end{array}               \right)
\end{eqnarray}
where $\tau_{\pm} = (1 \pm e^{-4 \kappa t}) / 2$. If one chooses the isotropic noise, 
Eq.(\ref{lindbald}) yields
\begin{eqnarray}
\label{mixed-I}
\varepsilon_I(\rho_{EPR}) = \frac{1}{2}
\left(        \begin{array}{cccc}
       \tilde{\tau}_+  &  0  &  0  &  2 \tilde{\tau}_+ - 1          \\
       0  &  \tilde{\tau}_-  &  0  &  0          \\
       0  &  0  &  \tilde{\tau}_-  &  0          \\
       2 \tilde{\tau}_+ - 1  &  0  &  0  & \tilde{\tau}_+
              \end{array}               \right)
\end{eqnarray}
where $\tilde{\tau}_{\pm} = (1 \pm e^{-8 \kappa t}) / 2$. 

To address the second issue we consider a square of fidelity between $\rho_{in}$ and 
$\rho_{out}$
\begin{equation}
\label{fidelity1}
F(\rho_{in}, \rho_{out}) = \langle \psi_{in} | \rho_{out} | \psi_{in} \rangle
\equiv F(\theta, \phi).
\end{equation}
Then how much information Alice can send to Bob with imperfect EPR resource
$\varepsilon(\rho_{EPR})$ can be measured by the average fidelity
\begin{equation}
\label{fidelity2}
\bar{F} \equiv \frac{1}{4 \pi} \int_{0}^{2 \pi} d \phi \int_{0}^{\pi} d \theta
\sin \theta F(\theta, \phi).
\end{equation}
Thus the perfect teleportation means $\bar{F} = 1$. Ref.\cite{oh02} has shown that for 
the same-axis noises the average fidelities become
\begin{equation}
\label{fidelity3}
\bar{F}_x = \bar{F}_y = \bar{F}_z = \frac{2}{3} + \frac{1}{3} e^{-4 \kappa t}
\end{equation}
while for the case of the isotropic noise $\bar{F}$ becomes
\begin{equation}
\label{fidelity4}
\bar{F}_I = \frac{1}{2} + \frac{1}{2} e^{-8 \kappa t}.
\end{equation}
Regardless of types of the noisy channels $\bar{F}$ decays as $\kappa t$ increases.

What kind of information on the average fidelity $\bar{F}$ can be obtained from the 
entanglement of the mixed states $\varepsilon_{\alpha} (\rho_{EPR})$ ($\alpha = x,y,z$)
and $\varepsilon_I(\rho_{EPR})$ or {\it vice versa}? To address this quuestion is the main
motivation of this paper. Since $\bar{F}$ decreases with increasing $\kappa t$, we
can conjecture that the effect of noises generally disentangles the mixed states 
provided the entanglement is genuine resource for the teleportation. Since, furthermore,
$\bar{F} = 2/3$ corresponds to the best possible score when Alice and Bob communicate
with each other through classical channel\cite{popescu94-1}, this fact implies that
$\varepsilon(\rho_{EPR})$ does not play any role as entanglement resource when 
$\bar{F} \leq 2/3$. Thus we can conjecture that $\varepsilon_{\alpha} (\rho_{EPR})$ 
($\alpha = x,y,z$) should be separable states as $\kappa t$ approaches to infinity
while $\varepsilon_I(\rho_{EPR})$ becomes separable when 
$\kappa t \geq \mu_{*} = (1/8) \ln 3$. If our conjecture is right, we can conjecture
$\bar{F}$ from the entanglement of the mixed-state resource without any calculation. 
Reversely, we can conjecture the entanglement of mixed states from the average fidelity.
This means that entanglement is genuine resource in the teleportation process even if 
noises are involved. Since explicit calculation of the $n$-qubit mixed-state 
entanglement is highly non-trivial when $n \geq 3$\footnote{For some entanglement 
measures it is also highly non-trivial to compute it even for $n = 2$.}, it may give 
valuable tool for the approximate conjecture of the entanglement. 

We will show that the above-mentioned conjectures on the relation between entanglement
of mixed-state and $\bar{F}$ are perfectly correct. This paper is organized as follows.
In section II we discuss the entanglement measures for the mixed states and their
inter-relations. It is found that not only the entanglement of formation but also 
the Groverian measure are monotonically related to the concurrence. This fact indicates
that the optimal ensemble for the concurrence is also optimal for the Groverian 
measure. In section III we compute explicitly the concurrence, entanglement of formation,
and Groverian measure for various mixed-states obtained by same-axis and isotropic
noises. The results of the computation are compared to the average fidelity $\bar{F}$.
It is shown that as we conjectured, all entanglement measures become zero when 
$\bar{F} \leq 2/3$. To confirm that our conjecture is right, we also compute the 
entanglement measures and average fidelity for different-axis noises in 
section IV. In these cases the results perfectly agree with our conjecture. In section V
the optimal decomposition for the higher-qubit mixed states is discussed. Especially,
the case of three-qubit mixed-state is discussed by adopting quantum teleportation with
W-state. Also the calculability for the second definition of the Groverian measure is 
briefly discussed in the same section.

\section{Entanglement of mixed-states}

There are many measures which quantify the entanglement of the mixed states. Among them 
we will use in this paper the entanglement of formation and the Groverian measure.

As we said in the previous section the entanglement of formation for any pure bipartite
system is defined as a von Neumann entropy of its subsystems. Then using a convex
roof construction\cite{benn96,uhlmann99-1}, one can extend the definition of the 
entanglement of formation to the full state space in a natural way as
\begin{equation}
\label{entform1}
{\cal E} (\rho) = \min \sum_j P_j {\cal E} (\rho_j)
\end{equation}
where minimum is taken over all possible ensembles of pure states $\rho_j$
with $0 \leq P_j \leq 1$. In Ref.\cite{form2,form3} it was shown how to construct the 
optimal ensemble, where the minimization in Eq.(\ref{entform1}) is naturally taken
in two-qubit system.

A convex roof method also can be used to extend the definition of the Groverian measure
in the full state space
\begin{equation}
\label{groverian1}
G (\rho) = \min \sum_j P_j G (\rho_j)
\end{equation}
where minimum is taken over all possible ensembles of pure states. Since the Groverian
measure for pure state is entanglement monotone\cite{vidal98-1}, it is not difficult 
to prove that $G (\rho)$ in Eq.(\ref{groverian1}) is also monotone even if 
$\rho$ is mixed state.

However, there is different extension of the Groverian measure from the aspect of the 
operational treatment of the entanglement\cite{shapira06}. In Ref.\cite{shapira06}
the Groverian measure for mixed state is defined as 
\begin{equation}
\label{groverian2}
\tilde{G}(\rho) = \sqrt{1 - \max_{\sigma \in {\cal S}} F^2(\rho, \sigma)}
\end{equation}
where ${\cal S}$ is a set of separable states and $F(\rho, \sigma)$ is a fidelity 
defined $F(\rho, \sigma) = \mbox{Tr} \sqrt{\rho^{1/2} \sigma \rho^{1/2}}$. 
It was shown in Ref.\cite{shapira06} that $\tilde{G}(\rho)$ is also entanglement 
monotone. 
Following Uhlmann theorem\cite{uhlmann76} one can re-express $\tilde{G}(\rho)$ in a 
form
\begin{equation}
\label{groverian3}
\tilde{G}(\rho) = \sqrt{1 - \max_{|\phi\rangle} \max_{|\psi\rangle} |\langle \phi | \psi
              \rangle |^2 }
\end{equation}
where $|\phi\rangle$ and $|\psi\rangle$ are purifications of $\sigma$ and $\rho$
respectively\footnote{In fact, one can remove the optimization on 
$|\psi\rangle$\cite{nielsen00}, which yields
$$
\tilde{G}(\rho) = \sqrt{1 - \max_{|\phi\rangle}  |\langle \phi | \psi
              \rangle |^2 }.
$$ }.

Now, we would like to comment how the optimization for the Groverian measure 
defined in Eq.(\ref{groverian1}) is taken. In order to describe this it is 
convenient to comment first how the optimization for the entanglement of formation
was taken in Ref.\cite{form2,form3}. Firstly, authors in these references notified that
in pure $2$-qubit state $|\psi\rangle$ the entanglement of formation 
${\cal E} (\psi)$ and concurrence ${\cal C} (\psi)$ are related to 
each other in a form
\begin{equation}
\label{relation2.1}
{\cal E} (\psi) = h \left( \frac{1 + \sqrt{1 - {\cal C}^2 (\psi)}   }{2}
                                      \right)
\end{equation}
where $h(x) \equiv -x \log_2 x - (1-x) \log_2 (1-x)$. Thus ${\cal E} ({\cal C})$ is 
monotonically increasing from $0$ to $1$ as ${\cal C}$ goes from $0$ to $1$. For
the mixed states, therefore, optimization for the concurrence in all possible pure-state
ensembles naturally coincides with optimization for the entanglement of formation. 
Secondly, authors in Ref.\cite{form2} found the optimization for the concurrence by 
making use of some geometrical argument when the density matrix has two or three
zero eigenvalues. Finally, Wootters derived the optimal ensemble for arbitrary 
two-qubit mixed states in Ref.\cite{form3}. We should note that the Groverian measure
for arbitrary two-qubit pure state $|\psi\rangle$ is related to the concurrence in a 
form
\begin{equation}
\label{relation2.2}
G(\psi) = \frac{1}{\sqrt{2}} \left(1 - \sqrt{1 - {\cal C}^2 (\psi)} \right)^{1/2}.
\end{equation}
Like the entanglement of formation, therefore, $G({\cal C})$ is also monotonic
function from $0$ to $1/\sqrt{2}$ as ${\cal C}$ goes from $0$ to $1$. This supports
that the optimization for the concurrence in all possible ensembles of pure states
coincides with not only that for the entanglement of formation but also that for
the Groverian measure defined in Eq.(\ref{groverian1}).

Although, therefore, the optimization for the first Groverian measure $G(\rho)$ is 
possible, the optimization for the second Groverian measure $\tilde{G}(\rho)$ seems to 
be highly non-trivial because it is defined by the Groverian measure for $4$-qubit
pure states via the purification and Uhlmann theorem. In this paper we will use 
${\cal E} (\rho)$ and $G (\rho)$ to confirm our conjecture on the relation between 
the mixed-state entanglement and the average fidelity $\bar{F}$.

\section{same-axis and isotropic noises}
In this section we would like to compute the entanglement for the mixed states given in
Eq.(\ref{mixed-xyz}) and Eq.(\ref{mixed-I}). Before starting computation it is convenient
for later use to introduce a ``magic basis''\cite{benn96}:
\begin{eqnarray}
\label{magic}
& & |e_1 \rangle = \frac{1}{\sqrt{2}} \left( |00\rangle + |11\rangle \right)
\hspace{1.0cm}
   |e_2 \rangle = \frac{i}{\sqrt{2}} \left( |00\rangle - |11\rangle \right)
                                                               \\  \nonumber
& &|e_3 \rangle = \frac{i}{\sqrt{2}} \left( |01\rangle + |10\rangle \right)
\hspace{1.0cm}
   |e_4 \rangle = \frac{1}{\sqrt{2}} \left( |01\rangle - |10\rangle \right).
\end{eqnarray}

Now let us consider ($L_{2,x}$, $L_{3,x}$) noise which makes the EPR resource as 
$\varepsilon_x (\rho_{EPR})$ in Eq.(\ref{mixed-xyz}). Since 
$\varepsilon_x (\rho_{EPR})$ has two zero eigenvalues, one can construct the optimal 
ensemble of pure states by two different ways explained in Ref.\cite{form2} and 
Ref.\cite{form3} respectively. It is not difficult to show that both methods yield same
optimal ensemble whose explicit expression is 
\begin{equation}
\label{optimal-x}
\varepsilon_x (\rho_{EPR}) = \sum_{i=1}^2 P_i |X_i\rangle \langle X_i|
\end{equation}
where $P_1 = P_2 = 1/2$ and 
\begin{equation}
\label{optimal-x2}
|X_1\rangle = \sqrt{\tau_+} |e_1\rangle + i \sqrt{\tau_-} |e_3\rangle
\hspace{1.0cm}
|X_2\rangle = \sqrt{\tau_+} |e_1\rangle - i \sqrt{\tau_-} |e_3\rangle.
\end{equation}
Since the concurrence for arbitrary 
$2$-qubit state $|\psi\rangle = \sum_{i=1}^4 \alpha_i |e_i\rangle$ 
is $|\sum_i \alpha_i^2 |$, $|X_1\rangle$ and $|X_2\rangle$ have same concurrence
\begin{equation}
\label{con-x}
{\cal C}_x = {\cal C} (|X_1\rangle) = {\cal C} (|X_2\rangle) = \tau_+ - \tau_- = 
                                                   e^{-4 \kappa t}.
\end{equation}

\begin{figure}[ht!]
\begin{center}
\includegraphics[height=10.0cm]{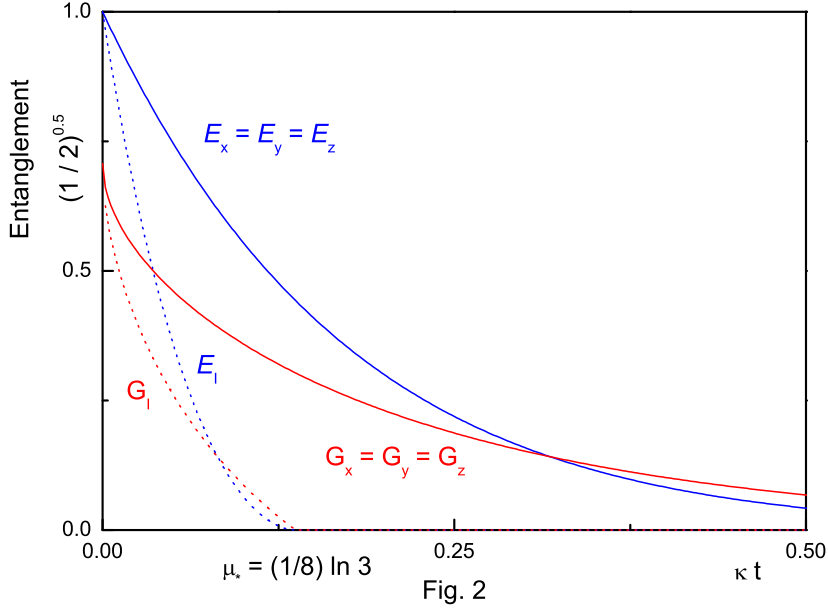}
\caption[fig2]{The $\kappa t$-dependence of the entanglement formation and 
Groverian measure for $\varepsilon_{\alpha} (\rho_{EPR})$ $(\alpha = x, y, z)$ and 
$\varepsilon_I (\rho_{EPR})$. Regardless of noise types the entanglement decreases with
increasing $\kappa t$. This means that the noises generally disentangle the quantum 
channel. For isotropic noisy channel ${\cal E}_I$ and $G_I$ become zero when
$\kappa t \geq \mu_* = (1/8) \ln 3$, where the average fidelity $\bar{F}$ is less
than $2/3$.}
\end{center}
\end{figure}

Thus the entanglement of formation ${\cal E}_x$ and the Groverian measure $G_x$ can be 
easily computed by Eq.(\ref{relation2.1}) and Eq.(\ref{relation2.2}) respectively. The 
$\kappa t$-dependence of ${\cal E}_x$ and $G_x$ are plotted in Fig. 2 as solid lines.
As expected ${\cal E}_x$ and $G_x$  decrease from $1$ and $1/\sqrt{2}$ to $0$ as 
$\kappa t$ goes from $0$ to $\infty$. This means that the noise disentangles  
$\varepsilon_x (\rho_{EPR})$ as we conjectured. Since ${\cal E}_x = G_x = 0$ at 
$\kappa t \rightarrow \infty$, $\varepsilon_x (\rho_{EPR})$ 
should be separable in this limit. 
We can confirm this directly from Eq.(\ref{optimal-x2}) because $|X_1\rangle$ and 
$|X_2\rangle$ reduce to 
$(|0\rangle \mp |1\rangle) / \sqrt{2} \otimes (|0\rangle \mp |1\rangle) / \sqrt{2}$
at $\kappa t \rightarrow \infty$ limit. If one constructs the optimal ensembles for 
$\varepsilon_y (\rho_{EPR})$ and $\varepsilon_z (\rho_{EPR})$, one can show by same 
way that $\varepsilon_y (\rho_{EPR}) = \sum_{i=1}^2 P_i |Y_i\rangle \langle Y_i|$ where 
$P_1 = P_2 = 1/2$ and 
\begin{equation}
\label{optimaly}
|Y_1\rangle = \sqrt{\tau_+} |e_1\rangle + i \sqrt{\tau_-} |e_4\rangle
\hspace{1.0cm}
|Y_2\rangle = \sqrt{\tau_+} |e_1\rangle - i \sqrt{\tau_-} |e_4\rangle,
\end{equation}
and $\varepsilon_z (\rho_{EPR}) = \sum_{i=1}^2 P_i |Z_i\rangle \langle Z_i|$ 
where $P_1 = P_2 = 1/2$
and 
\begin{equation}
\label{optimalz}
|Z_1\rangle = \sqrt{\tau_+} |e_1\rangle + i \sqrt{\tau_-} |e_2\rangle
\hspace{1.0cm}
|Z_2\rangle = \sqrt{\tau_+} |e_1\rangle - i \sqrt{\tau_-} |e_2\rangle.
\end{equation} 
It is easy
to show ${\cal E}_x = {\cal E}_y = {\cal E}_z$ and $G_x = G_y = G_z$.

Now, let us consider $\varepsilon_I (\rho_{EPR})$. Taking into account the partial 
transposition\cite{peres96,horod96,horod97} of $\varepsilon_I (\rho_{EPR})$ 
with respect to its subsystems, one
can realize that $\varepsilon_I (\rho_{EPR})$ is separable when 
$\kappa t \geq \mu_{*} = (1/8) \ln 3$. Following Ref.\cite{form3}, one can derive the 
separable decomposition 
$\varepsilon_I (\rho_{EPR}) = \sum_{i=1}^{4} |S_i\rangle \langle S_i|$ in this region,
where $|S_i\rangle$ are un-normalized vectors defined 
\begin{eqnarray}
\label{separable1}
& &|S_1\rangle = \frac{1}{2} \left( e^{i \theta_1} |x_1\rangle + e^{i \theta_2} 
|x_2\rangle + e^{i \theta_3} |x_3\rangle + e^{i \theta_4} |x_4\rangle \right)
                                                                    \\  \nonumber
& &|S_2\rangle = \frac{1}{2} \left( e^{i \theta_1} |x_1\rangle + e^{i \theta_2} 
|x_2\rangle - e^{i \theta_3} |x_3\rangle - e^{i \theta_4} |x_4\rangle \right)
                                                                    \\  \nonumber
& &|S_1\rangle = \frac{1}{2} \left( e^{i \theta_1} |x_1\rangle - e^{i \theta_2} 
|x_2\rangle + e^{i \theta_3} |x_3\rangle - e^{i \theta_4} |x_4\rangle \right)
                                                                    \\  \nonumber
& &|S_1\rangle = \frac{1}{2} \left( e^{i \theta_1} |x_1\rangle - e^{i \theta_2} 
|x_2\rangle - e^{i \theta_3} |x_3\rangle + e^{i \theta_4} |x_4\rangle \right).
\end{eqnarray}
In Eq.(\ref{separable1}) $|x_i\rangle$ are
\begin{eqnarray}
\label{separable2}
& &|x_1\rangle = -i \sqrt{\frac{3 \tilde{\tau}_+ - 1}{2}} |e_1\rangle
\hspace{1.0cm}
|x_2\rangle = -i \sqrt{\frac{\tilde{\tau}_-}{2}} |e_2\rangle
                                                              \\  \nonumber
& &|x_3\rangle = -i \sqrt{\frac{\tilde{\tau}_-}{2}} |e_3\rangle
\hspace{1.0cm}
|x_4\rangle = -i \sqrt{\frac{\tilde{\tau}_-}{2}} |e_4\rangle
\end{eqnarray}
and $\theta_i$'s satisfy
\begin{equation}
\label{separable3}
\frac{3 \tilde{\tau}_+ - 1}{\tilde{\tau}_-} e^{2 i \theta_1} + 
(e^{2 i \theta_2} + e^{2 i \theta_3} + e^{2 i \theta_4}) = 0.
\end{equation}
Since all $|S_i\rangle$ have zero concurrence provided Eq.(\ref{separable3}) holds, 
$\varepsilon_I (\rho_{EPR})$ becomes separable in the region $\kappa t \geq \mu_*$.
In order to see this explicitly let us consider the boundary of this region
$\kappa t = \mu_*$. At this point we have $\theta_1 = 0$ and 
$\theta_2 = \theta_3 = \theta_4 = \pi / 2$ which yield a following separable 
decomposition 
$\varepsilon_I (\rho_{EPR}) = \sum_{i=1}^4 P_i |\tilde{s}_i\rangle \langle \tilde{s}_i|$
where $P_1 = P_2 = P_3 = P_4 = 1/4$ and 
\begin{eqnarray}
\label{separable4}
& &|\tilde{s}_1\rangle = \left(\omega_- |0\rangle - \omega_+ e^{i \pi/4} |1\rangle\right)
\otimes \left(\omega_- |0\rangle - \omega_+ e^{-i \pi/4} |1\rangle\right)
                                                         \\   \nonumber
& &|\tilde{s}_2\rangle = \left(\omega_- |0\rangle + \omega_+ e^{i \pi/4} |1\rangle\right)
\otimes \left(\omega_- |0\rangle + \omega_+ e^{-i \pi/4} |1\rangle\right)
                                                         \\   \nonumber
& &|\tilde{s}_3\rangle = \left(\omega_+ |0\rangle - \omega_- e^{-i \pi/4} |1\rangle\right)
\otimes \left(\omega_+ |0\rangle - \omega_- e^{i \pi/4} |1\rangle\right)
                                                         \\   \nonumber
& &|\tilde{s}_4\rangle = \left(\omega_+ |0\rangle + \omega_- e^{-i \pi/4} |1\rangle\right)
\otimes \left(\omega_+ |0\rangle + \omega_- e^{i \pi/4} |1\rangle\right)
\end{eqnarray}                                                      
with $\omega_{\pm} = (\sqrt{3} (\sqrt{3} \pm 1) / 6)^{1/2}$.

In $\kappa t \leq \mu_*$ region $\varepsilon_I (\rho_{EPR})$ is generally entangled. 
The optimal ensemble of pure states can be constructed following Ref.\cite{form3}.
The final expression of decomposition is 
$\varepsilon_I (\rho_{EPR}) = \sum_{i=1}^4 P_i |I_i\rangle \langle I_i|$ where
$P_1 = P_2 = P_3 = P_4 = 1/4$ and
\begin{eqnarray}
\label{optimal-I}
& &|I_1\rangle = \sqrt{\lambda_1} |e_1\rangle - i \sqrt{3 \lambda_2} |e_2\rangle
                                                     \\   \nonumber
& &|I_2\rangle = \sqrt{\lambda_1} |e_1\rangle + i \sqrt{\frac{\lambda_2}{3}} |e_2\rangle
-2 i  \sqrt{\frac{2 \lambda_2}{3}} |e_3\rangle
                                                     \\  \nonumber
& &|I_3\rangle = \sqrt{\lambda_1} |e_1\rangle + i \sqrt{\frac{\lambda_2}{3}} |e_2\rangle
+ i \sqrt{\frac{2 \lambda_2}{3}} |e_3\rangle 
                                  -i \sqrt{2 \lambda_2} |e_4\rangle
                                                                 \\  \nonumber
& &|I_4\rangle = \sqrt{\lambda_1} |e_1\rangle + i \sqrt{\frac{\lambda_2}{3}} |e_2\rangle
+ i \sqrt{\frac{2 \lambda_2}{3}} |e_3\rangle 
                                  +i \sqrt{2 \lambda_2} |e_4\rangle
\end{eqnarray}
where $\lambda_1 = (3 \tilde{\tau}_+ - 1) / 2$ and $\lambda_2 = \tilde{\tau}_- / 2$.
It is easy to show that at the region $\kappa t \leq \mu_*$ 
$\varepsilon_I (\rho_{EPR})$ has a concurrence 
\begin{equation}
\label{con-I-1}
{\cal C}_I = \lambda_1 - 3 \lambda_2 = \frac{3}{2} \left(e^{-8 \kappa t} - \frac{1}{3}
                                                   \right).
\end{equation}
Since $\varepsilon_I (\rho_{EPR})$ is separable mixed state at $\kappa t \geq \mu_*$, 
${\cal C}_I$ equals to zero in this region. Thus we can write in a form
\begin{equation}
\label{con-I-2}
{\cal C}_I = \mbox{Max} \left(\lambda_1 - 3 \lambda_2, 0\right).
\end{equation}
Inserting Eq.(\ref{con-I-2}) into Eq.(\ref{relation2.1}) and Eq.(\ref{relation2.2}),
one can easily compute the entanglement of formation ${\cal E}_I$ and the 
Groverian measure $G_I$ for $\varepsilon_I (\rho_{EPR})$.

The $\kappa t$-dependence of ${\cal E}_I$ and $G_I$ are plotted in Fig. 2 as dotted lines.
As we conjectured in section 1, ${\cal E}_I$ and $G_I$ decrease from $1$ and $1/\sqrt{2}$
to $0$ as $\kappa t$ goes from $0$ to $\mu_*$. This means that when $\bar{F} \leq 2/3$,
$\varepsilon_I (\rho_{EPR})$ cannot play any role as a quantum channel. This fact
also indicates that the entanglement is a genuine resource for the quantum 
communication. In order to confirm that our conjecture is right, we will consider 
the different-axis noises in the next section.

\section{different-axis noises}

\begin{figure}[ht!]
\begin{center}
\includegraphics[height=10.0cm]{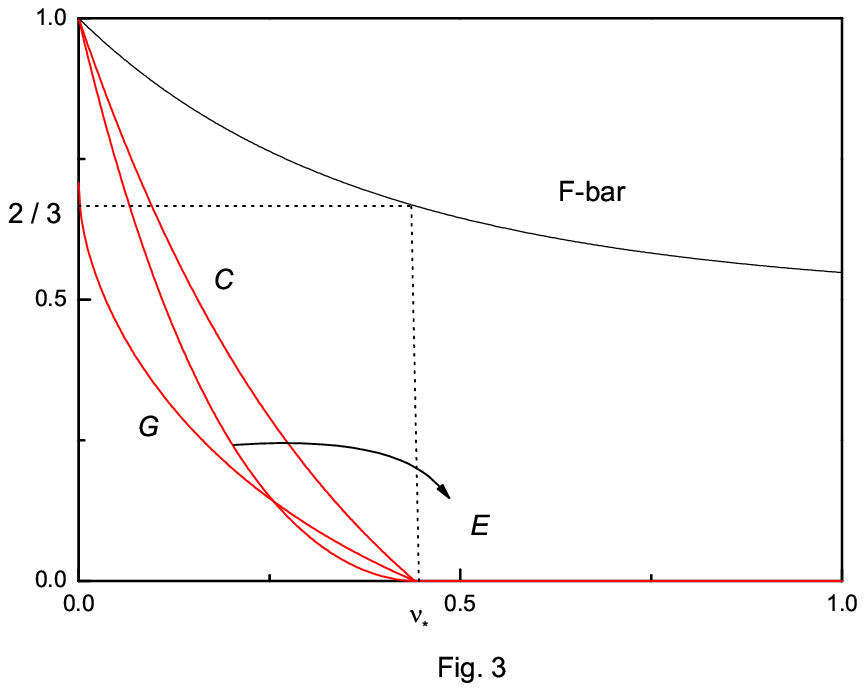}
\caption[fig3]{The $\kappa t$-dependence of the average fidelity $\bar{F}$, 
entanglement of formation ${\cal E}$, concurrence ${\cal C}$ and Groverian measure $G$
for different-axis noisy channels. As expected all entanglement measures reduce to 
zero when $\kappa t \geq \nu_* = \ln (1 + \sqrt{2}) / 2$. }
\end{center}
\end{figure}

In this section we would like to consider the different-axis noises to confirm that
our conjecture is right. First let us consider ($L_{2,x}$, $L_{3, z}$) noise. For 
this case the master equation (\ref{lindbald}) changes the EPR state $\rho_{EPR}$ into
\begin{eqnarray}
\label{mixed-xz1}
\varepsilon_{xz} (\rho_{EPR}) = \frac{1}{2}
\left(               \begin{array}{cccc}
       \nu_+  &  0  &  0  &  e^{-2 \kappa t} \nu_+    \\
       0  &  \nu_-  &  e^{-2 \kappa t} \nu_-  &  0    \\
       0  &  e^{-2 \kappa t} \nu_-  &  \nu_-  &  0    \\
       e^{-2 \kappa t} \nu_+  &  0  &  0  &  \nu_+
                      \end{array}                       \right)
\end{eqnarray}
where $\nu_{\pm} = (1 \pm e^{-2 \kappa t}) / 2$. Following the calculation of 
Ref.\cite{oh02}, one can show easily that the average fidelity in this noise channel
becomes 
\begin{equation}
\label{fidel-xz}
\bar{F} = \frac{1}{6} \left(3 + 2 e^{-2 \kappa t} + e^{-4 \kappa t} \right).
\end{equation}
Thus $\bar{F}$ becomes less than $2/3$ when 
$\kappa t \geq \nu_{*} = \ln (1 + \sqrt{2}) / 2$. We expect that 
$\varepsilon_{xz} (\rho_{EPR})$ becomes separable in the region $\kappa t \geq \nu_*$.
In fact, in this region $\varepsilon_{xz} (\rho_{EPR})$ can be expressed as 
$\varepsilon_{xz} (\rho_{EPR}) = \sum_{i=1}^4 |\bar{s}_i\rangle \langle \bar{s}_i|$
where $|\bar{s}_i\rangle$ are unnormalized vectors defined by same with 
Eq.(\ref{separable1}), but $|x_i\rangle$ are 
\begin{eqnarray}
\label{separable-xz1}
& &|x_1 \rangle = -i \nu_+ |e_1\rangle  \hspace{1.0cm}
|x_2\rangle = -i \sqrt{\nu_+ \nu_-} |e_2\rangle
                                           \\       \nonumber
& &|x_3\rangle = -i \sqrt{\nu_+ \nu_-} |e_3\rangle    \hspace{1.0cm}
|x_4\rangle = -i \nu_- |e_4\rangle
\end{eqnarray}
and $\theta_i$'s satisfy
\begin{equation}
\label{separable-xz2}
e^{2 i \theta_1} \nu_+^2 + \left(e^{2 i \theta_2} + e^{2 i \theta_3} \right) \nu_+ \nu_-
+ e^{2 i \theta_4} \nu_-^2 = 0.
\end{equation}
Since all $|\bar{s}_i\rangle$ have zero concurrence, $\varepsilon_{xz} (\rho_{EPR})$ is 
manifestly separable in $\kappa t \geq \nu_*$ as expected.

In the region $\kappa t \leq \nu_*$ we can derive an optimal ensemble of pure 
states. It needs a tedious calculation, and the final expression is 
$\varepsilon_{xz} (\rho_{EPR}) = \sum_{i=1}^4 P_i |XZ_i\rangle \langle XZ_i|$ where
$P_1 = P_2 = \nu_+ / (1 + 2 \nu_+)$, $P_3 = P_4 = 1 / (2 (1 + 2 \nu_+))$, and 
\begin{eqnarray}
\label{optimal-xz1}
& &|XZ_1\rangle = \nu_+ |e_1\rangle - i \sqrt{\nu_- (1 + \nu_+)} |e_2\rangle
                                                          \\  \nonumber
& &|XZ_2\rangle = \nu_+ |e_1\rangle + i \nu_+ \sqrt{\frac{\nu_-}{1 + \nu_+}} |e_2\rangle
- i \nu_+ 
\sqrt{\frac{\nu_- (1 + 2 \nu_+)}{1 + \nu_+}} |e_3\rangle
                                                          \\  \nonumber
& &|XZ_3\rangle = \nu_+ |e_1\rangle + i \nu_+ \sqrt{\frac{\nu_-}{1 + \nu_+}} |e_2\rangle
+ i \nu_+ \sqrt{\frac{\nu_- (1 + 2 \nu_+)}{1 + \nu_+}} |e_3\rangle
- i \nu_- \sqrt{1 + 2 \nu_+} |e_4\rangle
                                                         \\   \nonumber
& &|XZ_4\rangle = \nu_+ |e_1\rangle + i \nu_+ \sqrt{\frac{\nu_-}{1 + \nu_+}} |e_2\rangle
+ i \nu_+ \sqrt{\frac{\nu_- (1 + 2 \nu_+)}{1 + \nu_+}} |e_3\rangle
+ i \nu_- \sqrt{1 + 2 \nu_+} |e_4\rangle.
\end{eqnarray}
Using Eq.(\ref{optimal-xz1}) it is easy to compute the concurrence whose explicit 
expression is 
\begin{equation}
\label{con-xz}
{\cal C}_{xz}  = \frac{1}{2} \left( e^{-4 \kappa t} + 2 e^{-2 \kappa t} - 1\right)
\end{equation}
at $\kappa t \leq \nu_*$. Thus in the full range of $\kappa t$ ${\cal C} (\rho_{EPR})$
can be written as 
\begin{equation}
\label{con-diff}
{\cal C}_{xz}  = \mbox{Max} \left[ \frac{1}{2} \left( e^{-4 \kappa t} + 
                          2 e^{-2 \kappa t} - 1\right), 0 \right].
\end{equation}
Inserting Eq.(\ref{con-diff}) into Eq.(\ref{relation2.1}) and Eq.(\ref{relation2.2}), 
one can compute straightforwardly the entanglement of formation and the Groverian 
measure for $\varepsilon_{xz} (\rho_{EPR})$.

For ($L_{2,x}$, $L_{3,y}$) and ($L_{2,y}$, $L_{3,z}$) noises the EPR state becomes 
respectively
\begin{eqnarray}
\label{mixed-diff}
& &\varepsilon_{xy} (\rho_{EPR}) = \frac{1}{2} 
\left(              \begin{array}{cccc}
      \tau_+  &  0  &  0  &  e^{-2 \kappa t}   \\
      0  &  \tau_-  &  0  &  0                 \\
      0  &  0  &  \tau_-  &  0                 \\
      e^{-2 \kappa t}  &  0  &  0  &  \tau_+
                    \end{array}                       \right)
                                                             \\  \nonumber
& &\varepsilon_{yz} (\rho_{EPR}) = \frac{1}{2} 
\left(              \begin{array}{cccc}
          \nu_+  &  0  &  0  &  e^{-2 \kappa t} \nu_+    \\
          0  &  \nu_-  &  e^{-2 \kappa t} \nu_-  &  0    \\
          0  &  e^{-2 \kappa t} \nu_-  &  \nu_-  &  0    \\
          e^{-2 \kappa t} \nu_+  &  0  &  0  &  \nu_+  
                    \end{array}                       \right).
\end{eqnarray}
It is not difficult to show that the average fidelity for these are equal to 
Eq.(\ref{fidel-xz}) and their concurrences are same with Eq.(\ref{con-diff}), {\it i.e.}
concurrence for $\varepsilon_{xz} (\rho_{EPR})$. The optimal ensembles are 
$\varepsilon_{xy} (\rho_{EPR}) = \sum_{i=1}^4 P_i |XY_i\rangle \langle XY_i|$ and 
$\varepsilon_{yz} (\rho_{EPR}) = \sum_{i=1}^4 P_i |YZ_i\rangle \langle YZ_i|$, 
where $P_1 = P_2 = \nu_+ / (1 + 2 \nu_+)$ and $P_3 = P_4 = 1 / (2 (1 + 2 \nu_+))$.
The optimal pure states $|YZ_i\rangle$ can be obtained from $|XZ_i\rangle$ by 
interchanging $|e_3\rangle$ and $|e_4\rangle$. The optimal vectors $|XY_i\rangle$ are
obtained from $|XZ_i\rangle$ by cyclic change, {\it i.e.} 
$|e_2\rangle \rightarrow |e_3\rangle$, $|e_3\rangle \rightarrow |e_4\rangle$,
$|e_4\rangle \rightarrow |e_2\rangle$. The remaining different-axis noises 
($L_{2,z}$, $L_{3,x}$), ($L_{2,z}$, $L_{3,y}$), ($L_{2,y}$, $L_{3,x}$) generate similar
quantum channels to Eq.(\ref{mixed-xz1}) and Eq.(\ref{mixed-diff}). They also yield 
same average fidelities and same concurrences.

The average fidelity $\bar{F}$, concurrence ${\cal C}$, entanglement of formation
${\cal E}$ and the Groverian measure $G$ are plotted in Fig. 3. As expected, all 
entanglement meaures reduce to zero at $\kappa t \geq \nu_*$. Thus our conjecture 
described in section 1 is perfectly correct. This fact indicates that the entanglement
of the quantum channel is a genuine physical resource in the teleportation process.
Also our conjecture may offer valuable clues for the optimal decomposition in the 
higher-qubit mixed states. This will be discussed briefly in the next section.

\section{conclusion}

\begin{figure}[ht!]
\begin{center}
\includegraphics[height=10.0cm]{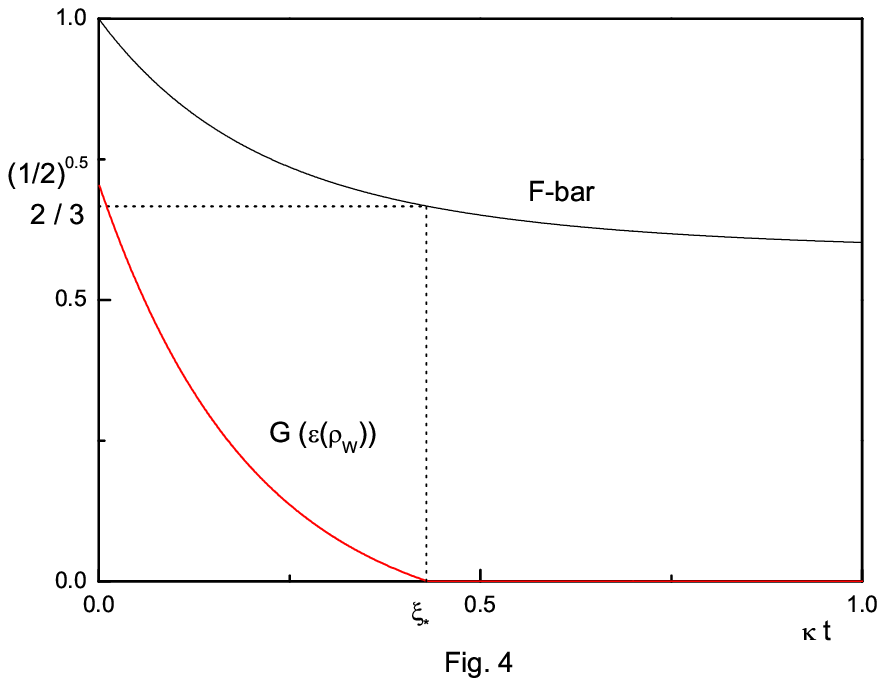}
\caption[fig4]{The $\kappa t$-dependence of the average fidelity $\bar{F}$ and 
Groverian measure $G$ for $3$-qubit mixed state $\varepsilon (\rho_W)$. The optimal
ensemble for $\varepsilon (\rho_W)$ should make $G$ to be zero 
when $\kappa t \geq \xi_*$. This may give valuable information for the construction of 
the optimal ensemble for higher-qubit states.}
\end{center}
\end{figure}

In this paper we have examined the connection between the mixed state entanglement and 
the average fidelity $\bar{F}$ using usual EPR-state teleportation via noises. As we
have shown, the mixed state entanglement becomes zero when $\bar{F} \leq 2/3$, which 
indicates that the entanglement of quantum channel is a genuine resource for 
teleportation. 

It is generally non-trivial task to compute the entanglement of $n$-qubit 
mixed states when $n \geq 3$. As far as we know, in addition, there is no way to 
find an optimal ensemble of pure states when $n \geq 3$. Also we cannnot define the
concurrence because there is no ``magic''-like basis in higher-qubit system. However, 
the result of our paper may provide a valuable information on the entanglement of 
higher-qubit mixed states. For example, let us consider $3$-qubit mixed state
\begin{eqnarray}
\label{qo-w2}
\varepsilon(\rho_W) = \frac{1}{16}
\left(           \begin{array}{cccccccc}
2 \alpha_2 & 0 & 0 & \sqrt{2} \alpha_2 & 0 & \sqrt{2} \alpha_2 & \alpha_2 & 0  \\
0 & 2 \alpha_1 & \sqrt{2} \alpha_1 & 0 & \sqrt{2} \alpha_1 & 0 & 0 & \alpha_3  \\
0 & \sqrt{2} \alpha_1 & 2 \beta_+ & 0 & \alpha_1 & 0 & 0 & \sqrt{2} \alpha_3  \\\sqrt{2} \alpha_2 & 0 & 0 & 2 \beta_- & 0 & \alpha_4 & \sqrt{2} \alpha_4 & 0   \\
0 & \sqrt{2} \alpha_1 & \alpha_1 & 0 & 2 \beta_+ & 0 & 0 & \sqrt{2} \alpha_3   \\
\sqrt{2} \alpha_2 & 0 & 0 & \alpha_4 & 0 & 2 \beta_- & \sqrt{2} \alpha_4 & 0   \\
\alpha_2 & 0 & 0 & \sqrt{2} \alpha_4 & 0 & \sqrt{2} \alpha_4 & 2 \alpha_4 & 0  \\
0 & \alpha_3 & \sqrt{2} \alpha_3 & 0 & \sqrt{2} \alpha_3 & 0 & 0 & 2 \alpha_3
                 \end{array}
                                                            \right)
\end{eqnarray}
where
\begin{eqnarray}
\label{def-w1}
& &\alpha_1 = 1 + e^{-2 \kappa t} + e^{-4 \kappa t} + e^{-6 \kappa t}  \\  \nonumber
& &\alpha_2 = 1 + e^{-2 \kappa t} - e^{-4 \kappa t} - e^{-6 \kappa t}  \\  \nonumber
& &\alpha_3 = 1 - e^{-2 \kappa t} - e^{-4 \kappa t} + e^{-6 \kappa t}  \\  \nonumber
& &\alpha_4 = 1 - e^{-2 \kappa t} + e^{-4 \kappa t} - e^{-6 \kappa t}  \\  \nonumber
& &\beta_{\pm} = 1 \pm e^{-6 \kappa t}.
\end{eqnarray}
This mixed state is constructed when the quantum teleportation is performed with W-state
\begin{equation}
\label{w-state}
|\psi_W\rangle = \frac{1}{2} \left( |100\rangle + |010\rangle + \sqrt{2} |001\rangle
                                                      \right)
\end{equation}
if ($L_{2,x}$, $L_{3,x}$, $L_{4,x}$) noise is introduced\cite{all1}. It has been shown
in Ref.\cite{all1} that its average fidelity between Alice and Bob is 
\begin{equation}
\label{w-fidel}
\bar{F} = \frac{1}{24} \left(14 + 3 e^{-2 \kappa t} + 2 e^{-4 \kappa t} + 5 
                             e^{-6 \kappa t} \right).
\end{equation}
Thus $\bar{F}$ decreases from $1$ to $7/12$ as $\kappa t$ goes from $0$ to $\infty$. 
From this fact we can conjecture that the Groverian measure (\ref{groverian1}) for 
$\varepsilon(\rho_W)$ decreases from $1/\sqrt{2}$ to $0$ when $\kappa t$ goes from $0$ to 
$\xi_* = 0.431041$ if we find the optimal ensemble of pure states for this mixed state.
This conjecture is described in Fig. 4. This information may give valuable clues for the 
construction of the optimal ensemble of pure states in three- or higher-qubit system.

Another point we would like to note is on the second definition of the Groverian measure
$\tilde{G}(\rho)$ defined in Eq.(\ref{groverian2}). Since it is not defined by convex
roof construction due to its operational meaning, we cannot use usual optimal ensemble 
technique to compute it. Since, furthermore, it is expressed as Eq.(\ref{groverian3}) 
via Uhlmann's theorem, we should know how to compute the Groverian measure of $n$-qubit
pure states with $n \geq 4$. Even if we assume that 
we have formula for $n$-qubit pure-state Groverian
measure, it is also highly non-trivial to take a maximization over all possible
purification. Since, however, it is a genuine entanglement measure for mixed states,
it should satisfy our conjecture. It may shed light on the development of the 
computational technique for $\tilde{G}(\rho)$ in the future.

{\bf Acknowledgement}: 
This work was supported by the Kyungnam University
Foundation Grant, 2008.

\end{document}